\documentclass[aps,pra,amsmath,amssymb,tightenlines,epsfig,floatfix,twocolumn,superscriptaddress, longbibliography]{revtex4-1}
\usepackage{graphicx}
\usepackage{dcolumn}	% Align table columns on decimal point
\usepackage{bm}			% bold math
\usepackage{amsfonts}
\usepackage{xspace}
\usepackage{color}
\usepackage{epstopdf}
\usepackage{multirow}

\begin{document}

\newcommand{\psihat}{\ensuremath{\hat{\psi}}\xspace}
\newcommand{\psihatd}{\ensuremath{\hat{\psi}^{\dagger}}\xspace}
\newcommand{\ahat}{\ensuremath{\hat{a}}\xspace}
\newcommand{\Ham}{\ensuremath{\mathcal{H}}\xspace}
\newcommand{\ahatd}{\ensuremath{\hat{a}^{\dagger}}\xspace}
\newcommand{\bhat}{\ensuremath{\hat{b}}\xspace}
\newcommand{\bhatd}{\ensuremath{\hat{b}^{\dagger}}\xspace}
\newcommand{\boldr}{\ensuremath{\mathbf{r}}\xspace}
\newcommand{\dr}{\ensuremath{\,d^3\mathbf{r}}\xspace}
\newcommand{\tr}{\ensuremath{\,\mathrm{Tr}}\xspace}
\newcommand{\dk}{\ensuremath{\,d^3\mathbf{k}}\xspace}
\newcommand{\etal}{\emph{et al.\/}\xspace}
\newcommand{\ie}{i.e.\ }
\newcommand{\eq}[1]{Eq.\ (\ref{#1})\xspace}
\newcommand{\fig}[1]{Fig.\ \ref{#1}\xspace}
\newcommand{\abs}[1]{\left| #1 \right|}
\newcommand{\proj}[2]{\left| #1 \rangle\langle #2\right| \xspace}
\newcommand{\Qhat}{\ensuremath{\hat{Q}}\xspace}
\newcommand{\Qhatd}{\ensuremath{\hat{Q}^\dag}\xspace}
\newcommand{\phihatd}{\ensuremath{\hat{\phi}^{\dagger}}\xspace}
\newcommand{\phihat}{\ensuremath{\hat{\phi}}\xspace}
\newcommand{\boldk}{\ensuremath{\mathbf{k}}\xspace}
\newcommand{\boldp}{\ensuremath{\mathbf{p}}\xspace}
\newcommand{\boldsigma}{\ensuremath{\boldsymbol\sigma}\xspace}
\newcommand{\boldalpha}{\ensuremath{\boldsymbol\alpha}\xspace}
\newcommand{\grad}{\ensuremath{\boldsymbol\nabla}\xspace}
\newcommand{\parti}[2]{\frac{ \partial #1}{\partial #2} \xspace}
 \newcommand{\vs}[1]{\ensuremath{\boldsymbol{#1}}\xspace}
\renewcommand{\v}[1]{\ensuremath{\mathbf{#1}}\xspace}
\newcommand{\Psihat}{\ensuremath{\hat{\Psi}}\xspace}
\newcommand{\Psihatd}{\ensuremath{\hat{\Psi}^{\dagger}}\xspace}
\newcommand{\Vhatd}{\ensuremath{\hat{V}^{\dagger}}\xspace}
\newcommand{\Xhat}{\ensuremath{\hat{X}}\xspace}
\newcommand{\Xhatd}{\ensuremath{\hat{X}^{\dag}}\xspace}
\newcommand{\Yhat}{\ensuremath{\hat{Y}}\xspace}
\newcommand{\Jhat}{\ensuremath{\hat{J}}\xspace}
\newcommand{\Yhatd}{\ensuremath{\hat{Y}^{\dag}}\xspace}
\newcommand{\Uhat}{\ensuremath{\hat{U}^{\dag}}\xspace}
\newcommand{\jhat}{\ensuremath{\hat{J}}\xspace}
\newcommand{\lhat}{\ensuremath{\hat{L}}\xspace}
\newcommand{\Nhat}{\ensuremath{\hat{N}}\xspace}
\newcommand{\rhohat}{\ensuremath{\hat{\rho}}\xspace}
\newcommand{\ddt}{\ensuremath{\frac{d}{dt}}\xspace}
\newcommand{\nset}{\ensuremath{n_1, n_2,\dots, n_k}\xspace}
\newcommand{\Var}{\ensuremath{\mathrm{Var}}\xspace}
\newcommand{\Erf}{\ensuremath{\mathrm{Erf}}\xspace}

\newcommand{\notes}[1]{{\color{blue}#1}}
\newcommand{\sah}[1]{{\color{magenta}#1}}
	
\title{Searching for Signatures of Quantum Gravity in Quantum Gases. }
\author{Simon A.~Haine}
\email{simon.a.haine@gmail.com}
\affiliation{Department of Quantum Science, Research School of Physics and Engineering, The Australian National University, Canberra, Australia}

\begin{abstract}
We explore the possibility of testing the quantum nature of the gravitational field with an ensemble of ultra-cold atoms. The use of many microscopic particles may circumvent some of the experimental obstacles encountered in recent proposals involving a pair of particles with mesoscopic mass. We employ multi-parameter estimation techniques, including the quantum and classical Fisher information to provide a criteria for the observability of the quantum effects, and compare to other recently proposed schemes. Crucially, we find that by preparing the appropriate initial state, interactions mediated via a quantum-valued gravitational field provide a signature that is distinct from classical gravitational interactions. We find that a test with ultra-cold atoms would be challenging, but not implausible with moderate improvements to current experimental techniques. 
\end{abstract}

\maketitle
While a full theoretical quantum treatment of quantum gravity remains elusive, \emph{effective} quantum field theories of gravity have been developed, by treating the quantum field as a perturbation from the classical solution \cite{Donoghue:1994, Burgess:2004, Carney:2019}. Low-energy laboratory-scale experiments are well within this perturbative regime \cite{Carney:2019}. However, whether or not the the gravitational field displays quantum properties, that is, if a quantum treatment is required at all, is a question that demands experimental resolution \cite{Feynman:1957}, with several proposed theories stating that the gravitational field may be fundamentally classical \cite{Diosi:1984, Penrose:1996, Kafri:2014, Kafri:2015}. There have been several recent proposals for laboratory scale experiments to discriminate between these theories and a theory with a quantised gravitational field \cite{Kafri:2014, Kafri:2015, Bahrami:2015, Anastopoulos:2015, Tilloy:2016, Gan:2016, Derakhshani:2016, Carney:2019}, and experiments that rule out classes of semi-classical gravity \cite{Page:1981, Altamirano:2018}. Recently, it has been argued that if the gravitational interaction between two quantum systems can create entanglement, then the gravitational field must act as a quantum mediator \cite{Kafri:2014, Kafri:2015, Moulick:2016, Marletto:2017c, Marletto:2017b}. This principle is the basis of recent experimental proposals for testing this phenomenon by searching for gravitational induced entanglement between two freely-falling micron-sized test masses \cite{Bose:2017, Marletto:2017}, and more recently, optomechanical systems \cite{Carlesso:2019, Krisnanda:2019}. While these proposals are feasible with a modest improvement in current experimental capabilities, there are still significant experimental challenges associated with them. As such, it is worth exploring if the differences that other systems provide can prove advantageous. 

In this paper, we investigate the gravitational interaction amongst an ensemble of ultra-cold atoms. Ultra-cold atoms potentially provide a number of advantages over other systems when performing ultra-precise measurements. Of particular relevance is that the response of an atom to electromagnetic and gravitational fields is fundamentally locked to physical constants, which provides a long-term stability which no macroscopic system can. This property differentiates atoms from quantum systems formed from a particular macroscopic objects, such as a nano-diamond or a micro-mechanical oscillator, where the mass or spring-constant may have a slight drift over the duration required to collect sufficient experimental statistics. This is of particular importance when searching for a weak effect that will require a large number of repetitions, and is the reason why atoms are the system of choice for atomic clocks and ultra-precise gravimeters \cite{Cronin:2009, Kritsotakis:2018}. 

Bose \etal \cite{Bose:2017} and Marletto and Vedral \cite{Marletto:2017} have previously shown that quantum entanglement between two systems which is generated by the gravitational field is indirect evidence of quantum gravity. Here, we take a slightly different approach, which is to consider how the effect of a quantum valued gravitational field would differ from a classical valued field. We note that in our approach, the presence of a quantum-valued field results in a Hamiltonian that can generate entanglement, while the Hamiltonian resulting from a classical-valued gravitational field cannot, implying logical consistency of our approach with the entanglement witness approaches. 
We find that if the gravitational interaction is mediated by a quantum-valued gravitational field, it presents different signatures to that of a classical-valued gravitational field. By carefully engineering the many-particle quantum state of the system, we can maximise the effect of the observability of these signatures. Other than the use of many particles, an important difference between this work and the work of Bose \etal \cite{Bose:2017} and Marletto and Vedral \cite{Marletto:2017} is that instead of searching for an entanglement witness, we formulate the experiment as a multi-parameter estimation problem, which, in principle allows us to search for arbitrarily weak interactions by performing many repetitions of the experiment. Crucially, we show that classical theories of gravity, and decoherence effects, produce a signatures in the measured probability distribution that are distinct from what an interaction mediated by a quantum-valued gravitational field predicts.  

\section{Gravitational interaction for an ensemble of two-level atoms}
We assume an ensemble of ultracold bosonic atoms with two electronic states $|a\rangle$ and $|b\rangle$. Following \cite{Bose:2017, Marletto:2017}, rather than attempting to present a quantum description of the gravitational field, we describe the gravitational interactions between the particles directly. The contribution to the Hamiltonian due to the gravitational interaction energy is given by
\begin{equation}
\hat{H}_G = -\frac{1}{2}G m^2 \int \int : \hat{n}(\boldr)\hat{n}(\boldr^\prime): \frac{1}{|\boldr - \boldr^\prime|} \dr \dr^\prime \label{ham1} ,
\end{equation}
where $m$ is the mass of each individual particle, and $G$ is Newtons constant for universal gravitation. The operator $\hat{n}$ is given by
\begin{equation}
\hat{n}(\boldr) = \psihatd_a(\boldr)\psihat_a(\boldr) + \psihatd_b(\boldr)\psihat_b(\boldr) \, ,
\end{equation}
and represents the total particle density, and  $\psihat_j(\boldr)$ is the standard bosonic field operator annihilating a particle from state $|j\rangle$ at point $\boldr$, obeying the standard commutation relations
\begin{equation}
\left[\psihat_i(\boldr) \, , \psihatd_j(\boldr^\prime)\right] = \delta_{ij}\delta(\boldr-\boldr^\prime) \, ,
\end{equation}
 and `$::$' denotes normal ordering. We have ignored the post-newtonian correction and first quantum correction, as for our system they are of orders $10^{-40}$ and $10^{-57}$ smaller \cite{Carney:2019}. It is important to note that in writing \eq{ham1}, we have implicitly assumed that the gravitational field can be quantum valued. Assuming that the interaction between the atoms is mediated by a gravitational field (i.e., not a direct non-local action at a distance), this field must be quantum valued. Specifically, \eq{ham1} can be re-written as
\begin{equation}
\hat{H}_G = \frac{m}{2} \int : \hat{n}(\boldr)\hat{\Phi}_G(\boldr) : \dr ,
\end{equation}
where 
\begin{equation}
\hat{\Phi}_G(\boldr) = -G m \int \hat{n}(\boldr^\prime) \frac{1}{|\boldr - \boldr^\prime|} \dr^\prime ,
\end{equation}
is the (operator valued) gravitational potential due to source masses at location $\boldr^\prime$, and the factor of $\frac{1}{2}$ removes double-counting. That is, the the gravitational potential is quantum valued as a result of the quantum uncertainty in the location of the source masses. 

Consider, in contrast, a \emph{classical}-valued gravitational potential $\Phi_C(\boldr)$, originating from some classical source. Placing our quantum test-particles in this field results in the gravitational potential
\begin{equation}
\hat{H}_C =  m \int \hat{n}(\boldr) \Phi_C(\boldr) \dr \label{hamc} .
\end{equation} 
Clearly, this gravitational interaction energy can also be quantum valued, but only due to the quantum-ness of the location of the particles, rather than the quantum-valued nature of the gravitational source (and therefore field) as in \eq{ham1}. 
A theory of gravity that does not allow for a quantum valued field would then be described by a Hamiltonian such as \eq{hamc}, with the value of $\Phi_C(\boldr)$ depending on how a classical valued field arrises from a quantum distribution of source masses. For example, in a semi-classical theory of gravity \cite{Rosenfeld:1963, Diosi:1984, Yang:2013, Gan:2016}, 
\begin{equation}
\Phi_C(\boldr) = \langle \hat{\Phi}_G(\boldr)\rangle .
\end{equation}
However, we do not restrict ourself to this case, and simply consider \emph{any} classical gravitational field, without making any assumptions on how the value of $\Phi_C(\boldr)$ is determined from the quantum distribution of source masses. Our task is then to devise an experiment that can distinguish between the effects of \eq{ham1} and \eq{hamc}. 

It has been shown that quantum entanglement can only be generated via a quantum field \cite{Mari:2016, Bose:2017, Marletto:2017, Belenchia:2018}. We note that the Hamiltonian \eq{ham1} is capable of generating entanglement between spatially separated particles, while \eq{hamc} can only perform local operations. Therefore, although we are not explicitly searching for the presence of entanglement as the signature of quantum gravity, distinguishing between the effects of two Hamiltonians, one which is capable of generating entanglement, and one which is not, is logically equivalent.

\begin{figure}
\includegraphics[width =0.8\columnwidth]{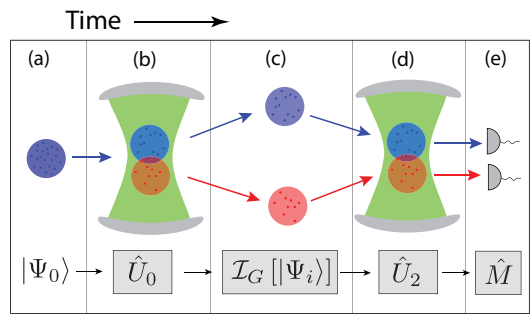}
\caption{Scheme for probing the gravitational interaction with ultra-cold atoms. (a): An ensemble of two level atoms are all prepared in state $|a\rangle$. (b): A state preparation unitary operation $\hat{U}_0$ distributes the atoms between two modes, and may create non-trivial quantum correlations. (c): The two modes are spatially separated and evolve for time $t$ under the gravitational interaction $\mathcal{I}_G$. (d): A second unitary operation $\hat{U}_2$ interferes the two modes, before the number in each mode is measured (e).  }
\label{fig:scheme}
\end{figure}

\section{Scheme to Search for Signatures of Quantum Gravity with Ultra-cold Atoms}
Our scheme is based on a standard Mach-Zehnder interferometer, and is illustrated in \fig{fig:scheme}. Each atom is initially prepared in state $|a\rangle$, before being placed in an equal coherent superposition of $|a\rangle$ and $|b\rangle$ at time $t=0$ via a $\pi/2$ pulse implemented via a radio-frequency transition. The two components are then spatially separated via a state dependent potential. Alternatively, if a two-photon Raman transition is used instead of a radio-frequency transition, the need for a potential is removed as the two components will separate due to the momentum transfer of the Raman transition \cite{Moy:1997, Hagley:1999}, which would remove any systematic error due to slight uncertainties in the behaviour of the potential. After evolving under the effects of the gravitational interaction for time $t$, the two components are then brought back together and recombined by a second $\pi/2$ pulse \footnote{If a Raman transition is used, an additional $\pi$ pulse would be needed at time $t/2$ in order to reverse the momentum. However, this does not effect the dynamics of the quantum state.}, and the number in each component is measured. Quantitatively, we describe the evolution as follows: The state $|\Psi_i\rangle = \hat{U}_0|\Psi_0\rangle$ is prepared by splitting the initial state into two components. We also allow for the possibility that $\hat{U}_0$ may contribute additional operations during this step, such as creating entanglement between the particles. The atoms interact gravitationally, via the interaction $\mathcal{I}_G\left[|\Psi_i\rangle\right]$, which may be a unitary mapping, or perhaps some operation that causes decoherence, before a second unitary operation $\hat{U}_2$ is applied, and a measurement $\hat{M}$ is made. 

If the atoms in state $|j\rangle$ remain localised in spatial mode $u_j(\boldr,t)$, then its reasonable to make the approximation
 \begin{equation}
 \psihat_j(\boldr) = \hat{j}u_j(\boldr,t)
 \end{equation}
 with
 \begin{equation}
 \left[\hat{j} , \hat{k}^\dag \right] = \delta_{jk}, \quad \left[\hat{j}^\dag , \hat{k}^\dag \right] = \left[\hat{j} , \hat{k} \right] =0
 \end{equation}
 Inserting this into \eq{ham1} gives
 \begin{equation}
 \hat{H} = \kappa_{aa}(t) \ahatd\ahatd\ahat\ahat + \kappa_{bb}(t) \bhatd\bhatd\bhat\bhat + 2\kappa_{ab}(t)\ahatd\ahat\bhatd\bhat \label{ham2}
 \end{equation}
 where 
 \begin{equation}
 \kappa_{ij} = -\frac{1}{2} G m^2\int \int |u_i(\boldr)|^2 |u_j(\boldr^\prime)|^2 \frac{1}{|\boldr - \boldr^\prime|} \dr \dr^\prime \, . \label{kappa_def}
 \end{equation}
 Meanwhile, making the same substitution in \eq{hamc} gives
 \begin{equation}
 \hat{H}_C = v_a \ahatd\ahat + v_b\bhatd\bhat \label{ham2c}
 \end{equation}
 where
 \begin{equation}
 v_j = m\int |u_j(\boldr)|^2 \Phi_c(\boldr) \dr \,.
 \end{equation}
If the wavepackets remain spatially separated by significantly more than their spatial extent for the majority of the evolution, $\kappa_{ab} \ll \kappa_{aa}, \kappa_{bb}$. This means that the gravitational potential energy of the system will be larger when all of the atoms are in one mode compared to the case when they are evenly distributed across both modes (and therefore occupying a larger volume). Setting $\kappa_{aa}= \kappa_{bb} = \kappa_0$, $\kappa_{ab}=0$,  the unitary operator for the evolution given by \eq{ham2} is
 \begin{equation}
 \hat{U}_Q = \exp\left(i \alpha \hat{A} \right)  \, . \label{UQ}
 \end{equation} 
where $\alpha = \kappa_0 t/\hbar$, and $\hat{A} = \ahatd\ahatd\ahat\ahat + \bhatd\bhatd\bhat\bhat$. In this case, the gravitational interaction is simply described by $\mathcal{I}_G\left[|\Psi_i\rangle\right] = U_Q|\Psi_i\rangle$.

\begin{figure*}
\includegraphics[width =\textwidth]{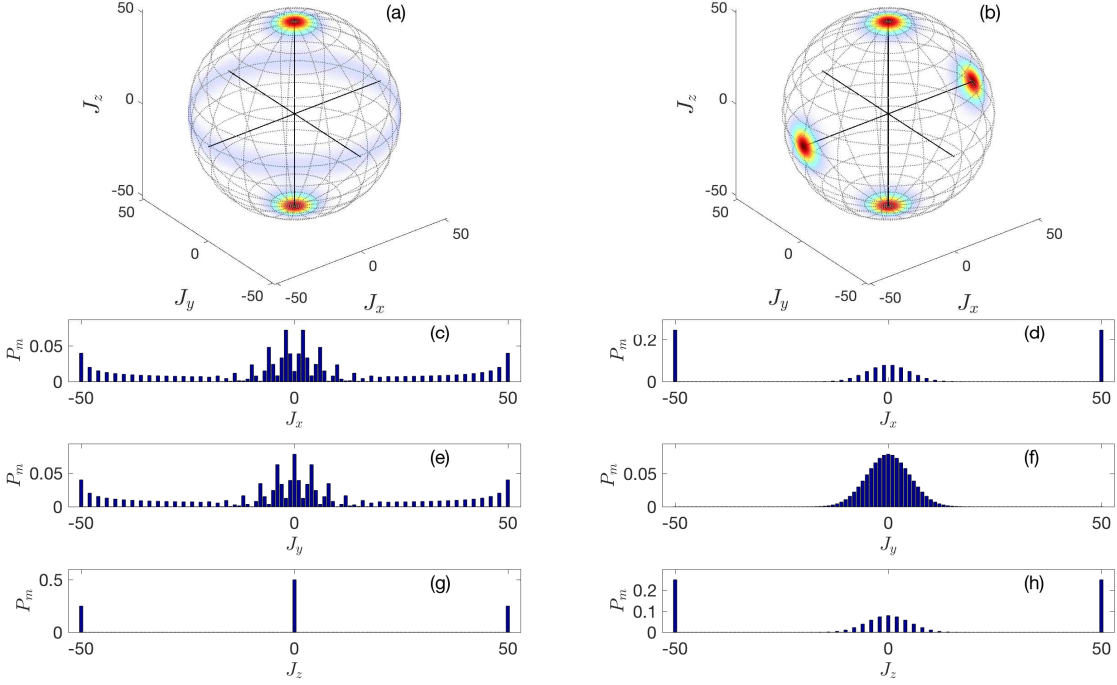}
\caption{The Husimi-$Q$ function $Q(\theta,\phi)$ for the state given in \eq{psi_opt} (a) and \eq{psi_cat} (b). The Husimi-$Q$ function is defined as  $Q(\theta, \phi) = |\langle \xi(\theta, \phi) |\Psi\rangle |^2$, where $|\xi(\theta,\phi)\rangle = \exp(i\phi\hat{J}_z) \exp(i\theta\hat{J}_x)|(N/2)_z\rangle$ is a coherent spin state. The projection into the $\jhat_x$, $\jhat_y$ and $\jhat_z$ basis are also shown for \eq{psi_opt} (c, e, g) and \eq{psi_cat} (d, f, h) respectively. }
\label{statefig1}
\end{figure*}

Meanwhile, the unitary evolution operator resulting from \eq{ham2} is 
 \begin{equation}
 \hat{U}_C= \exp\left[i (\beta\jhat_z  + \gamma \hat{N}_+)\right]
 \end{equation}
 where $\beta = (v_b-v_a)t$, $\gamma = \frac{-1}{2}(v_a+v_b)t$, $\hat{N}_+ = \ahatd \ahat + \bhatd\bhat$, and $\jhat_z = \frac{1}{2}(\ahatd\ahat - \bhatd\bhat)$, and $\mathcal{I}_G\left[|\Psi_i\rangle\right] = U_C|\Psi_i\rangle$. If we assume a mean-field gravity model, then $\Phi(\boldr) = \langle \hat{\Phi}(\boldr)\rangle$, and we can estimate the expected values of $v_a$ and $v_b$. However, we do not restrict ourself to this case, and treat these as unknown parameters to allow for the possibility of other models of classical gravity. Our task of demonstrating the quantum valued nature of the gravitational field can now be summarised as inferring a non-zero value of $\alpha$, while distinguishing this from the effects of a possibly non-zero, unknown value of $\beta$ and $\gamma$.  Fortunately, we can ignore $\gamma$, as for a fixed total number of particles, this parameter results only in a  global phase shift to the state, which is undetectable. We therefore estimate the smallest value of $\alpha$ that can be detected in the presence of the unknown \emph{nuisance} parameter $\beta$. We can place a bound on our ability to determine the value of $\alpha$ by introducing the quantum Fisher covariance matrix \cite{Paris:2009}, with elements 
\begin{equation}
F_{i,j} = 2\left(\langle \partial_i \Psi | \partial_j \Psi \rangle + \langle \partial_j \Psi | \partial_i \Psi \rangle - 2\langle \Psi | \partial_i \Psi \rangle \langle \partial_j \Psi | \Psi \rangle\right) \, . 
\end{equation}
for $\{i,j\} = \{\alpha,\beta\}$. 
In general, the quantum Fisher covariance matrix resulting from the state
\begin{equation}
|\Psi \rangle = \hat{U}_Q \hat{U}_C |\Psi_i\rangle \, ,
\end{equation}
contains off-diagonal terms, and therefore our ability to distinguish the presence of $\alpha$ from the possible presence of $\beta$ is hindered.  However, by carefully choosing our initial state, we can find a diagonal quantum Fisher covariance matrix with large diagonal terms. Introducing the collective spin operators $\jhat_k = \frac{1}{2}\bold{a}^\dag \sigma_k \bold{a}$, where $\bold{a} =  (\ahat, \bhat)^T$ and $\sigma_k$ is the $k$th Pauli matrix, and denoting $|m_k\rangle$ as the eigenstate of $\jhat_k$ with eigenvalue $m_k$, we find that the state
\begin{equation}
|\Psi_i\rangle = \frac{1}{\sqrt{2}}\left[|0_z \rangle + \frac{1}{\sqrt{2}}\left(|(N/2)_z \rangle + |(-N/2)_z \rangle \right)\right] \label{psi_opt} \, ,
\end{equation}
leads to a quantum Fisher covariance matrix with elements $F_{\alpha, \alpha} = N^4/4$, $F_{\beta, \beta} = 2 N^2$, and $F_{\alpha, \beta} = F_{\beta, \alpha} = 0$. This state is shown in \fig{statefig1}. The lack off-diagonal terms means that the matrix is trivial to invert, and we should be able to infer the presence of a non-zero $\alpha$ as long as
\begin{equation}
\alpha \gtrsim \frac{1}{\sqrt{k F_{\alpha,\alpha}}} = \frac{2}{\sqrt{k}N^2} \, , \label{delta_alpha}
\end{equation}
where $k$ is the number of repetitions of the experiment. We note that as this state is a superposition of maximal and minimal eigenstates of $\hat{A}$, it is the state that maximises $F_{\alpha, \alpha}$ for $N$ particles. 

\begin{figure} 
\includegraphics[width =\columnwidth]{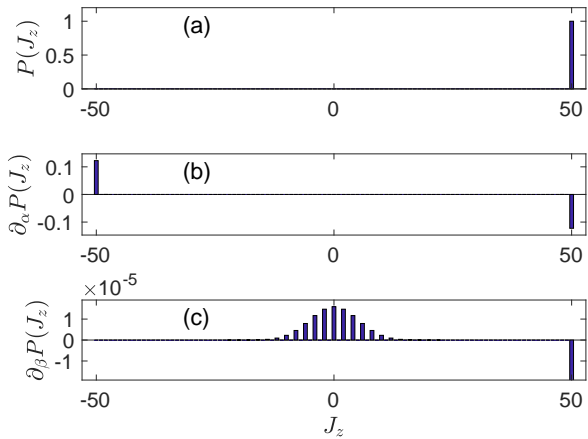}
\caption{(a) The projection into the $\jhat_z$ basis, $P(J_z)$ for the state $|\Psi_f\rangle = \hat{U}_0^\dag \hat{U}_Q \hat{U}_C \hat{U}_0|\Psi_0\rangle$, for $\alpha = 0$, $\beta = 0$. (b) $\partial_\alpha P(J_z)$ for $\alpha = 10^{-8} \ll 1/\sqrt{F_{\alpha,\alpha}}$. (c) $\partial_\beta P(J_z)$ for $\beta = 10^{-8} \ll 1/\sqrt{F_{\beta,\beta}}$. $N=100$ for all frames.}
\label{probfig1}
\end{figure}

The question remains how to engineer this state, and to find a measurement scheme that saturates the Quantum Cramer-Rao bound (QCRB) \cite{Braunstein:1994, Paris:2009, Toth:2014, Demkowicz-Dobrzanski:2014}. While standard atom-optics techniques cannot easily produce \eq{psi_opt}, we can create a state that performs nearly as well via the well-studied one-axis twisting (OAT) technique \cite{Kitagawa:1993, Sorensen:2001, Gross:2010, Riedel:2010, Leroux:2010, Haine:2014, Haine:2018, Nolan:2018}. In particular, we envisage a controllable one-axis twisting interaction via an optically induced nonlinearity as demonstrated in \cite{Leroux:2010, Hosten:2016b, Davis:2016, Borregaard:2017}. Starting with all the particles in mode $|a\rangle$, ie, $|\Psi_0\rangle = (\ahatd)^N/\sqrt{N!}|0\rangle = |(N/2)_z\rangle$, and applying $\pi/2$ rotation about the $\jhat_x$ axis, followed by a one-axis twisting interaction for period $\tau$, followed by a second $\pi/2$ rotation results in the state preparation unitary
\begin{equation}
\hat{U}_0 = \exp(i\jhat_x \frac{\pi}{2})\exp(i \chi \tau \jhat_z^2)\exp(i\jhat_x \frac{\pi}{2}) \, .
\end{equation}
Setting $\chi \tau = \frac{\pi}{4}$ gives the state shown in \fig{statefig1}b, which is 
\begin{equation}
|\Psi_i\rangle = \hat{U}_0 |\Psi_0\rangle = \mathcal{N} e^{i\frac{N}{2} \pi} \left(|\eta_x\rangle +  e^{-i\frac{3\pi}{4}} |\eta_z\rangle\right) \, , \label{psi_cat}
\end{equation}
where 
\begin{subequations}
\begin{align}
|\eta_x\rangle &= \frac{1}{\sqrt{2}}\left(|(N/2)_x\rangle + |(-N/2)_x\rangle\right) , \\
|\eta_z\rangle &= \frac{1}{\sqrt{2}}\left(|(N/2)_z \rangle - e^{i\frac{N}{2} \pi} |(-N/2)_z\rangle\right) , 
\end{align}
\end{subequations}
and the normalisation factor $\mathcal{N} \approx \frac{1}{\sqrt{2}}$ as a result of $|\langle \eta_x|\eta_z\rangle| \ll 1$ for  $N\gg 1$. This state has $F_{\alpha, \alpha} = N^4/4 - \mathcal{O}(N^3)$, $F_{\beta, \beta} = 2(N^2 + N)$, and  $F_{\alpha, \beta} =0$. However, the task remains to find a measurement that saturates the QCRB. That is, the obtainable sensitivity from a particular measurement is given by the diagonal terms of the covariance matrix, which can be calculated from the inverse of the \emph{classical} Fisher information matrix
\begin{equation}
\Delta \alpha^2 = \left[\mathcal{F}^{-1}\right]_{\alpha, \alpha} \, ,
\end{equation}
where $\mathcal{F}$ is the classical Fisher information matrix with elements
\begin{equation}
\mathcal{F}_{ij} = \sum_m \frac{\partial_i P_m \partial_j P_m}{P_m} .
\end{equation}
Following \cite{Macri:2016, Nolan:2017b, Mirkhalaf:2018}, we can saturate the QCRB by applying the unitary operator $\hat{U}_2 = \hat{U}_0^\dag$ after the interrogation time, and then measuring in a basis that projects into $|\Psi_0\rangle$, which can be achieved by measuring in the $\jhat_z$ basis. This results in a classical Fisher covariance matrix with elements $\mathcal{F}_{ij} = F_{ij}$, and therefore we can achieve the sensitivity given by \eq{delta_alpha}. To illustrate how the effect of two potential models of gravity can be distinguished, figure \ref{probfig1} shows $\partial_k P_{m_z}$ for $k = \alpha$, $\beta$. A small, non-zero value of $\alpha$ results in transfer of probability from $P_{N/2}$ to $P_{-N/2}$, while a small, non-zero value of $\beta$ results in transfer to a distribution of values around $m_z \approx 0$.  

\section{Effects of de-coherence}
Another potential effect that our measurement must be able to discriminate is the possibility that the gravitational field causes decoherence \cite{Diosi:1987, Penrose:1996, Kafri:2014, Kafri:2015, Altamirano:2018}. That is, superpositions of different gravitational source configurations suffer de-phasing.  We must also allow for the possibility of other unaccounted sources of decoherence, as this could cause a signature in the detection signal that is falsely attributed to a non-zero value of $\alpha$. We account for the possibility of decoherence (due to gravitational or other effects) phenomenologically by introducing an additional parameter or unknown magnitude $\delta_\Gamma$, such that the effect of the de-phasing interaction on our initial state is $\mathcal{I}_G\left[\hat{\rho}_i\right] = \exp(\delta_\Gamma \mathcal{L}[\Gamma])\hat{\rho}_i \equiv \hat{\rho} $, where 
\begin{align}
\mathcal{L}[\hat{\Gamma}]\hat{\rho}_i &= \hat{\Gamma} \hat{\rho}_i\hat{\Gamma} - \frac{1}{2}\left(\hat{\Gamma}^2 \hat{\rho}_i + \hat{\rho}\hat{\Gamma}^2 \right) \, ,
\end{align}
such that
\begin{align}
\hat{\rho}  = \sum_n\sum_m \exp(- \delta_\Gamma (\lambda_n-\lambda_m)^2) \proj{n}{n} \hat{\rho}_i  \proj{m}{m} \, ,
\end{align}
where $\hat{\rho}_i =  |\Psi_i\rangle\langle \Psi_i|$, and $\hat{\Gamma}|n\rangle = \lambda_n|n\rangle$ \cite{Wiseman:2010, Nolan:2017, Haine:2015b}. By setting $\hat{\Gamma} = \hat{A}$, the eigenstates represent states with well-defined gravitational interaction energy. That is, off-diagonal terms representing superpositions of states with a large difference in their gravitational interaction energy will suffer greater decoherence. Another possibility is that it is just superpositions of the \emph{source} potential, rather than the total interaction energy that decohere, which could be accounted for by setting $\hat{\Gamma} = \hat{J}_z$. This choice of $\hat{\Gamma}$ would also describe decoherence due to the presence of a fluctuating field interacting with a non-vanishing difference in the magnetic dipole moment of the two atomic species.

\begin{figure}
\includegraphics[width =\columnwidth]{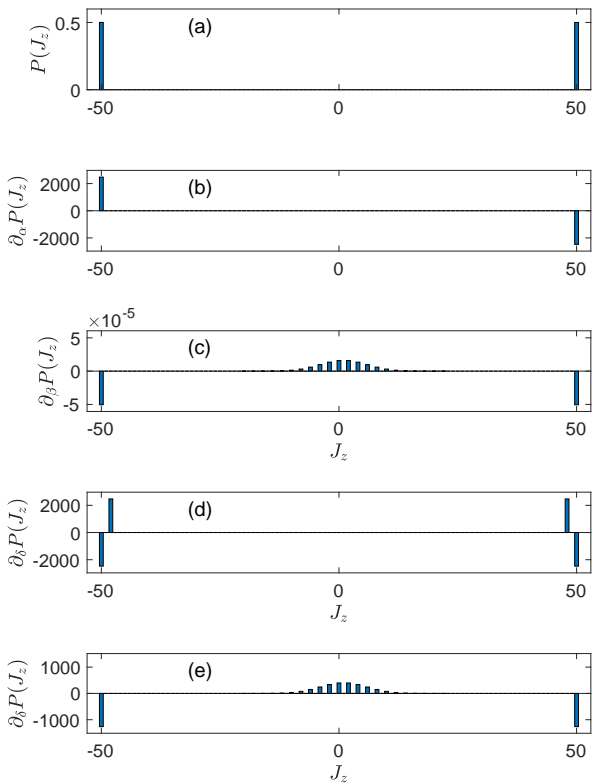}
\caption{(a) The projection into the $\jhat_z$ basis, $P(J_z)$ for the state $|\Psi_f\rangle = U_0 U_Q U_C U_0|\Psi_0\rangle$, for $\alpha = 0$, $\beta = 0$. (b) $\partial_\alpha P(J_z)$ for $\alpha = 10^{-8} \ll 1/\sqrt{F_{\alpha,\alpha}}$. (c) $\partial_\beta P(J_z)$ for $\beta = 10^{-8} \ll 1/\sqrt{F_{\beta,\beta}}$. (d) $\partial_{\delta_A} P(J_z)$ for $\hat{\Gamma} = \hat{A}$, and $\delta_A = 10^{-8}$. (e) $\partial_{\delta_{J_z}} P(J_z)$ for $\hat{\Gamma} = \hat{J_z}$, and $\delta_{J_z} = 10^{-8}$ ($N=100$ for all frames).}
\label{probfig2}
\end{figure}

While it is challenging to compute the full quantum Fisher covariance matrix including $\delta_A$ and $\delta_{J_z}$, it is straight-forward to numerically compute the classical Fisher covariance matrix. Unfortunately, the state preparation and measurement procedure presented in the previous section ($\hat{U}_2 = \hat{U}_0^\dag$, followed by measurement in the $\hat{J}_z$ basis) leads to a non-diagonal Fisher covariance matrix for the parameters $\alpha$, $\beta$, $\delta_A$ and $\delta_{J_z}$. In fact, setting $\hat{\Gamma} = \hat{A}$, results in $\partial_{\delta_A} P(J_z)$ with nearly identical profile to $\partial_\alpha P(J_z)$. Fortunately, an alternate choice of $\hat{U}_2$ remedies the situation.   Specifically, choosing $\hat{U}_2 = \hat{U}_0$ (rather than $\hat{U}_0^\dag$) and computing the Fisher covariance matrix numerically results in values of $\mathcal{F}_{\alpha, k} \ll \mathcal{F}_{\alpha, \alpha}$ for  $k \neq \alpha$, and as such can be inverted to find $\left[\mathcal{F}\right]^{-1}_{\alpha,\alpha} \approx F_{\alpha,\alpha}$. In figure \ref{probfig2} we show $P(J_z)$ and $\partial_k P(J_z)$ for $k = \alpha$, $\beta$, $\delta_{A}$, and $\delta_{J_z}$. Again, we see that $\beta$, $\delta_{A}$, and $\delta_{J_z}$ give signatures that are easy to distinguish from $\hat{U}_Q$. An additional benefit of using $\hat{U}_2 = \hat{U}_0$ (rather than $\hat{U}_2 = \hat{U}_0^\dag$) is that the final state has a maximally bi-modal probability distribution and is therefore optimally robust to detection noise such that our inference of $\alpha$ saturates the noisey quantum Cramer-Rao bound, as introduced in \cite{Haine:2018b}.

\section{Experimental considerations}
We can estimate the magnitude of $\alpha$ by assuming $u_{a(b)}(\boldr)$ are Gaussian wave-packets of the form
\begin{equation}
|u_{a(b)}(\boldr)|^2 = \frac{1}{\sigma^3 \pi^{3/2}}\exp\left[\frac{-(\boldr-(+) x_0 \mathbf{\hat{x}})^2}{\sigma^2}\right] \, .
\end{equation}
Inserting this into \eq{kappa_def} gives
\begin{equation}
\alpha = \frac{t G m^2}{\hbar \sigma \sqrt{\pi}} \, ,
\end{equation}
where we have assumed that $x_0 \gg \sigma$ such that we can ignore the contribution from $\kappa_{ab}$. Using \eq{delta_alpha}, the minimum number of particles required to observe signatures of quantum gravity is then
\begin{equation}
N = \sqrt{\frac{2\hbar \sigma \sqrt{\pi}}{\sqrt{k}G m^2 t}} \, .
\end{equation}
For realistic experimental parameters, ($\sigma =50\mu$m, $t=1$s, $k=10^5$) and considering $^{87}$Rb, the required number of particles is $N \approx 5 \times 10^{9}$, which is 2 orders of magnitude more than are used in current state-of-the-art precision sensing experiments. These parameters result in an atomic density of $\sim 4\times 10^{13}$ cm$^{-3}$, which is considerably less than the density in the precision atom-interferometry experiment reported in \cite{Altin:2011}. If we consider more massive atoms such as Ytterbium,  and the possibility of employing di-atomic molecules, then the required number of particles is reduced further. Additionally, increasing the interrogation time ($t$), or the number of repetitions ($k$) are also not inconceivable. The number of repetitions $k=10^5$ equates to approximately 1 day of continuous operation (with $t=1$s), or 10 days including $10$s of state preparation time. This duration of data collection is not uncommon in atomic squeezing experiments, such as \cite{Gross:2010}, for example. Although the preparation of exotic quantum states such as \eq{psi_cat} has not been demonstrated in large ensembles of particles, the field is currently in a state of rapid progress \cite{Hosten:2016, Engelsen:2017, Davis:2018, Lewis-Swan:2018}.   

Obviously, when searching for such minuscule signals, it is important to consider sources of systematic error. Linear shifts in energy such as Zeeman and Stark shifts due to fluctuating electric and magnetic fields, as well as fluctuations in any confining potentials could potentially be orders of magnitude larger than the gravitational signature. However, these shifts are all single-particle effects, and as our multi-parameter analysis reveals, with the appropriate choice of quantum state, these effects de-couple from the gravitational signature that we are searching for. That is, these effects would couple into the value of $\beta$ or $\delta_{J_z}$, which cause a change in the measured probability distribution that is entirely distinguishable from $\alpha$.  A potentially much more serious issue is imperfect knowledge of the inter-particle interactions due to short-range van der Waals interaction. For ultra-cold quantum gases, these interactions are dominated by $s$-wave scattering, which can be manipulated via a Feshbach resonance \cite{Pethick:2008}, or compensated for by an optically-induced atom-atom interaction \cite{Davis:2018}. However, any imperfect knowledge of the magnitude of this interaction will contribute to a term in \eq{ham2} that has the same form as the gravitational interaction. That is, they contribute to the coefficient of $\hat{A}$ and therefore are indistinguishable from a change in $\alpha$. To successfully perform this experiment at the precision required to observe gravitational effects, the electromagnetic interaction will need to be measured to much higher precision than the current state-of-the-art \cite{Egorov:2013}. Interestingly, the scheme provided in this paper may be capable of a very sensitive measurement of the scattering-length, and thus removing this source of systematic error. Alternatively, as this coefficient has different scaling with $\sigma$ compared to $\alpha$. That is, as gravity is a long-range force, $\alpha \propto \sigma^{-1}$. However, the magnitude of $s$-wave interactions that generally dominate ultra-cold atomic gases scale as $\sigma^{-3}$, so by varying the value of $\sigma$ it may be possible to isolate the two contributions. That is, doubling $\sigma$ reduces $\alpha$ by a factor of $2$, but will decrease the effect of the atomic interactions by a factor of $8$. 

\section{Discussion}
It is worth discussing how a scheme involving the use of many particles differs from the schemes proposed by Bose \etal \cite{Bose:2017}, and Marletto and Vedral \cite{Marletto:2017}, and how these differences might prove advantageous. Firstly, the use of many particles allows for a far greater space of possible quantum states, and therefore a much greater capacity to tailor the specific state to the particular requirements. Formulating the problem in the language of multi-parameter estimation, rather than searching for an entanglement witness, means that, in principle, arbitrary small signals can be found by performing more repetitions of the experiment to increase precision, assuming long-term stability can be achieved. Additionally, the freedom in the choice of the quantum state means that we can engineer our quantum state such that it is maximally sensitive to the parameter of interest (in our case, the gravitational interaction), and ensuring that this parameter provides a signature that is distinct from other effects. It is interesting to consider the scaling of the scheme with the number of particles, $N$. Increasing $N$ not only increases the magnitude of gravitational interaction, it also increases the ability of the quantum state to perceive a small change. Although we are not searching for entanglement, and therefore not directly demonstrating entanglement mediated through the gravitational field, a Hamiltonian of the form \eq{ham2} \emph{can} create entanglement, so proving that the gravitational field obeys a Hamiltonian of this form is evidence that the gravitational field can transmit entanglement.  It was recently argued by Hall and Reginatto that, depending on how quantum-classical interactions are modelled, the generation of entanglement through a gravitational interaction is \emph{not} definitive proof of non-classical gravity \cite{Hall:2018, Reginatto:2018}. It's possible that the scheme considered here could perhaps differentiate between a broader class of classical gravitational models than the schemes presented in Bose \etal \cite{Bose:2017} and Marletto and Vedral \cite{Marletto:2017}, however, further investigation is required. 

It is worth discussing how this proposal differs from the experimental results discussed in Altamirano \etal \cite{Altamirano:2018}. By considering  recent high-precision atom-interferometry experiments, they show that the observed results are inconsistent with gravity interacting as a local operations and classical communication (LOCC) channel, as proposed by Kafri, Taylor and Milburn \cite{Kafri:2014, Kafri:2015}, which should cause far greater de-coherence. This decoherence is the result of a single atom with quantum degrees of freedom interacting with the earth's gravitational field, and does not directly show that the gravitational field can generate entanglement between two spatially separated masses. The experiment that we propose would provide evidence that the gravitational field can exist in a quantum superposition.  

In summary, we have investigated a scheme to search for quantum signatures of gravity in ultra-cold quantum gases. The use of atoms rather than quantum systems formed from particular macroscopic objects provide much greater long-term stability, a factor which will be important when searching for such small signatures. Additionally,  the use of many microscopic particles and multi-parameter estimation may provide additional advantages over the previously proposed schemes.  Although it appears that the observation of quantum signatures of gravitation in ultra-cold quantum gases is beyond the ability of current experiments, improvements in particle number, and knowledge of the residual electro-magnetic forces between the atoms may enable the detection of effects of quantum gravity.

\vspace{0.5cm}
%\begin{acknowledgements}
\section*{acknowledgements}
The author acknowledges fruitful discussions with Joe Hope, Antigonie Bradshaw, Matthew Blacker, Ruvi Lecamwasam, John Close, Nick Robins, and Michael Hall. 
%The Australian National University is situated on land traditionally owned by the Ngunnawal people. 
%\end{acknowledgements}

%\bibliography{simon_bib.bib}
\bibliography{../../simon_bib.bib}

\end{document}